\documentclass[%
reprint,
superscriptaddress,
amsmath,amssymb,
aps,prb
]{revtex4-2}

\usepackage{graphicx}
\usepackage{bm}
\usepackage[dvipsnames]{xcolor}

\begin{document}

\title{Giant effective $g$-factor due to spin bifurcations in polariton condensates}

\author{A. Bochin}
\affiliation{School of Physics and Engineering, ITMO University, Kronverksky Pr. 49, bldg. A, St. Petersburg, 197101, Russia}
\author{I. Chestnov}
\affiliation{School of Physics and Engineering, ITMO University, Kronverksky Pr. 49, bldg. A, St. Petersburg, 197101, Russia}
\author{A. Nalitov}
\affiliation{Moscow Institute of Physics and Technology, Institutskiy per.,~9, Dolgoprudnyi, Moscow Region, Russia 141701}

\begin{abstract}
We predict giant susceptibility of spin-bifurcating polariton condensates to externally applied permanent magnetic field.
In the presence of spin-anisotropic polariton-polariton interactions, the condensate spontaneously takes an elliptically polarised state, {whose perturbation dynamics can be interpreted in terms of the presence of  strong effective magnetic field} significantly surpassing the external one. 
Surprisingly, this behaviour of the addressed strongly out-of-equilibrium system in the vicinity of a critical point exhibits intriguing analogy with the second-order phase transition. 
The predicted field-enhancement effect can be utilized for creation of  topologically nontrivial states of Bogoliubov's excitations existing on top of the polariton condensate.
\end{abstract}

\maketitle

Magneto-optic phenomena lifting degeneracy of photonic modes in the presence of external magnetic field are remarkable manifestations of light-matter coupling.
As an example, Zeeman effect for electron excitations in dielectrics or semiconductors causes energy splitting of otherwise degenerate circularly polarised photonic modes, which, in turn, results in magneto-optic Faraday effect or Kerr rotation \cite{born_principles_2019}.
The possibility of local time-reversal symmetry breaking for electromagnetic waves due to magneto-optic coupling is crucial for engineering topological photonic states and suppressing backscattering in signal transmission \cite{Haldane2008,ozawa2019}.

Although the magneto-optic effect is weak in the optical frequency domain, it can be significantly enhanced in the strong coupling regime, where mixed light-matter quasi-particles, such as exciton-polaritons, emerge.
In particular, topological exciton-polariton states were demonstrated in optical cavity lattices subject to strong magnetic fields \cite{Klembt2018}.
In addition, incorporating ferromagnetic materials in strongly coupled optical cavities was proposed recently to reach giant values of the effective $g$-factor, which quantifies the magneto-optic coupling strength \cite{lyons2022}.

The full potential of strongly coupled systems supporting exciton polaritons is revealed in the nonlinear regime, where the macroscopic coherent states associated with bosonic condensates are formed. 
In particular, spin-anisotropic polariton-polariton interaction results in Larmor precession of the condensate pseudospin in a self-induced effective magnetic field even in the absence of an externally applied one \cite{Laussy2006}.

In the thermodynamic equilibrium limit, the externally applied field and the effective self-induced one are opposite and even exactly compensate each other {below a certain threshold}, giving rise to the spin Meissner effect \cite{Rubo2006}.
In contrast, nonequilibrium condensates can develop spin polarisation and thus produce effective magnetic fields spontaneously due to the spin bifurcation mechanism \cite{Ohadi2015}. 
{Similarly, coherently driven nonequilibrium polariton states also exhibit spontaneous spin polarisation and spin multistability \cite{Gippius2007}.} {Note that the interplay between the equilibrium effect of locked polarisation direction in the spin Meissner regime and a spontaneous choice of spin polarisation typical to driven-dissipative condensates is still under discussion \cite{Krol2019,Sawicki2024}}.

Artificial gauge \cite{whittaker_data_2020} and Zeeman \cite{Rechtsman2013} fields emerging in spatially structured optical systems can replace the external magnetic field giving rise to topological photonic states.
Similarly, the effective field in spontaneously spin-polarised condensate lattices was shown to result in topologically nontrivial excitation spectra and unidirectional edge states \cite{Sigurdsson2019}.
However, the underlying spin bifurcation mechanism requires a delicate balance and degeneracy of the spin states, rendering this system extremely susceptible to external symmetry breaking factors.
This inspires investigation of the possibility to exploit this sensitivity to control the band topological invariants in polariton condensate lattices.

In this work, we show that the effective field of a single spin-bifurcating condensate is not only aligned with the external Zeeman field, but can also significantly surpass it in magnitude.
Magnification of the weak Zeeman field with the strong and aligned effective field can be considered as a giant enhancement of the effective $g$-factor.
We demonstrate that the proposed enhancement is most pronounced in the vicinity of the critical bifurcation point and identify the range of the optimal parameters.

The system of interest represents a condensate of exciton-polaritons whose coherent state is governed by the driven-dissipative Gross-Pitaevskii equation for a spinor wave function $\mathbf{\Psi} = \left(\Psi_+,\Psi_-\right)^\intercal$:{
\begin{eqnarray}
    i\dot{\Psi}_{\pm} = &-&\frac{i}{2} \left(i\Delta + W - \Gamma - \eta N_{\rm pol} \right)\Psi_{\pm} - 
    \frac{1}{2}\left(\varepsilon + i\gamma \right)\Psi_{\mp}  \notag 
    \\
    &+& \frac{1}{2} \left( \alpha_1 \left|\Psi_{\pm} \right|^2 + \alpha_2\left|\Psi_{\mp} \right|^2 \right)\Psi_{\pm},
\end{eqnarray}
where} $\Psi_{+(-)}$ stands for spin-up (down) projections of the polariton state on the structure growth axis {corresponding to the right (left) circular polarisation of the emitted photons}. Here $\Gamma$ is the polariton decay rate, $W$ is the rate of stimulated scattering from the incoherent particle reservoir, $\eta$ {is} the strength {of the gain saturation proportional to the condensate occupation $N_{\rm pol}$}, while $\varepsilon$ and $\gamma$ are respectively the energy splitting and the loss-rate difference of the two linearly polarised polariton states. {Finally, $\alpha_{1(2)}$ is the interaction constant for polaritons with the same (opposite) spin.}

Following Ref.~\cite{Ohadi2015}, we treat the system using the classical pseudospin vector $\mathbf{S} = 1/2 \mathbf{\Psi}^{\dag}\bm{\sigma}\mathbf{\Psi}$ defined via Pauli vector $\bm{\sigma}$ with the magnitude $S$ proportional to the condensate occupation and the direction characterising its spin state.
{Note that the pseudospin $\mathbf{S}$ is directly connected to the Stokes parameters of the laser emission as $S = S_0$, $S_x = S_1$, $S_y = S_2$, $S_z = S_3$.}
The pseudospin dynamics is described with the following system of equations:
\begin{subequations}  \label{eq:Sxyz}
\begin{eqnarray} 
    \dot{S}_x &=& (W - \eta S - \Gamma) S_x - \gamma S - (\alpha S_z + \Delta) S_y, \\
    \dot{S}_y &=& (W - \eta S - \Gamma) S_y + \varepsilon S_z + (\alpha S_z + \Delta) S_x, \\
    \dot{S}_z &=& (W - \eta S - \Gamma) S_z - \varepsilon S_y.
\end{eqnarray}
\end{subequations}

In analogy with the classical spin in magnetic field, the pseudospin vector $\mathbf{S}$ {is precessing} about the $z$-axis due to the mixed effect of Zeeman splitting $\Delta$ from the external out-of-plane magnetic field and a self-induced effective field $\alpha S_z$ \cite{Shelykh2004,Laussy2006}. 
The latter is governed by the interaction constant $\alpha {=\alpha_1-\alpha_2} > 0$, which accounts for {the strong spin anisotropy of polariton-polariton} interactions.

In what follows, we use the dimensionless form of system (\ref{eq:Sxyz}):
\begin{subequations} \label{eq:sxyz} 
\begin{eqnarray} 
    \dot{s}_x &=& (p-s)s_x - g s - (a s_z + \delta) s_y, \\
    \dot{s}_y &=& (p-s)s_y + s_z + (a s_z + \delta) s_x, \\
    \dot{s}_z &=& (p-s)s_z - s_y,
\end{eqnarray}
\end{subequations}
with the effective splitting field $\delta = \Delta/\varepsilon$, the dimensionless time $\tau = t\varepsilon$ and a couple of dimensionless parameters $a = \alpha / \eta$, $g = \gamma / \varepsilon$. 
We also define the effective pump strength $p = (W-\Gamma)/\varepsilon$ and the normalized Stokes vector $\mathbf{s} = \eta \mathbf{S}/\varepsilon$.

In the absence of an external magnetic field, $\delta = 0$, a couple of linearly polarised ($s_z = s_y = 0$) trivial solutions of system \eqref{eq:sxyz} are characterised by the magnitudes $s_\pm(p) = p\pm g$.
These states are pinned to the $x$-axis, $s_x = \mp s_\pm^{(0)}$.
At $g>0$ (the upper linearly polarised mode dissipates stronger than the lower one), the $s_-$-state is dynamically unstable against weak perturbations according to the regular Lyapunov stability analysis.
The more populated state $s_+$ is stable below the critical pumping $p_c = (1-ag+g^2)/a$, characterised by the condensate population $s_c = (g^2+1)/a$.

\begin{figure*}
\centering
    \includegraphics[width = 0.75\textwidth]{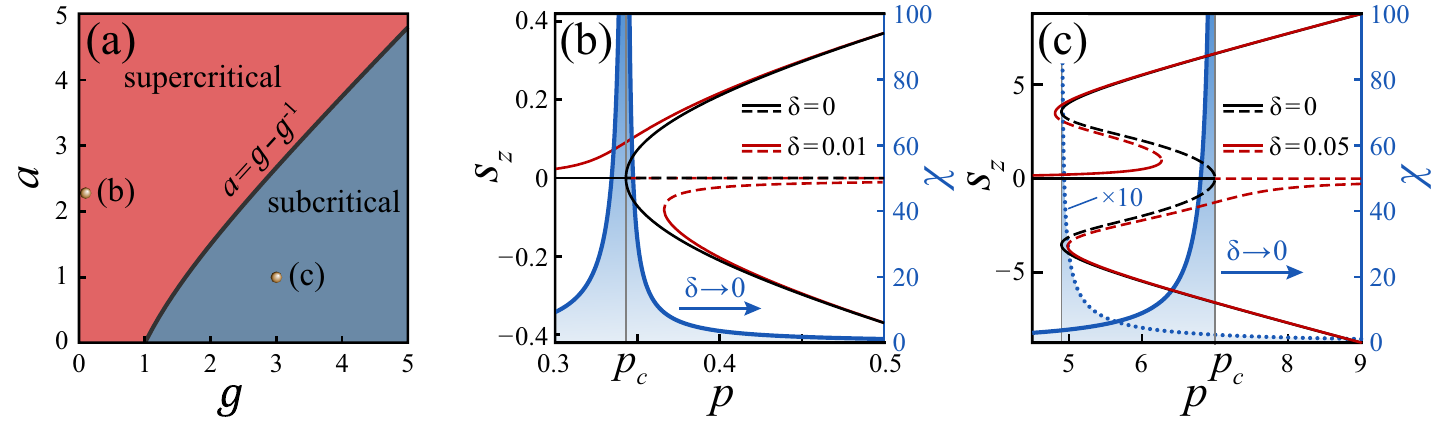}
    \caption{\textbf{Fig. 1.} (a) The phase diagram of the spontaneous symmetry breaking pitchfork bifurcation at $\delta=0$. Pseudospin projection $s_{z}$ (left axis) and field enhancement parameter $\chi$ (right axis) in the supercritical regime at $g=0.1$ and $a = 2.28$ \cite{Ohadi2015}. The solid lines correspond to stable states while dashing shows dynamically unstable solutions. (c) The same as on panel (b) but in the subcritical regime at $g=3$ and $a=1$. The dotted blue curve corresponds to the field-enhancement $\chi$ of the upper stable elliptically polarised state which appears in the fold bifurcation.}
    \label{Fig1}
\end{figure*}

At $p=p_c$, the $s_+$-state is destabilized via a pitchfork bifurcation, which has a supercritical character if $a>g-g^{-1}$ and a subcritical one otherwise (see the phase diagram in Fig.~\ref{Fig1}(a)).
This gives rise to a couple of nontrivial elliptically polarised ($s_z \neq 0$) states with
\begin{subequations} \label{eq:s0el}
\begin{eqnarray} \label{eq:s0xy}
    s_x = (p-s)s/g, \:\ \ \  s_y = (p-s) s_z,\\
    s_z = \pm {s \over g} \sqrt{g^2 - (p-s)^2 \over 1 + (p-s)^2}, \label{eq:s0z}
\end{eqnarray}
\end{subequations}
and the condensate population $s$ given by the positive root of the quadratic equation
\begin{equation}
    \left( {a \over g} - 1 \right) s^2 - \left({a \over g} - 2 \right) p s - (p^2 + 1) = 0.
\end{equation}

Note that at $g<0$, the symmetry-breaking bifurcation does not alter stability properties of $s_{\pm}$-states.
In this regime, the more populated state $s_-$ always remains the only stable linearly polarised solution.
In what follows, we focus on the case $g>0$, corresponding to experimental conditions \cite{Ohadi2015}.


To describe the impact of the external magnetic field, we start with the limit of perturbatively weak~field~$\delta$.
In the first approximation, stationary solutions acquire a correction $\mathbf{s^{(1)}} = \left( s_x^{(1)}, s_y^{(1)}, s_z^{(1)}\right)^\intercal$ linear in $\delta$:
\begin{equation} \label{eq:corr1}
    \mathbf{s}^{(1)} = \delta \times \left[J\left(\mathbf{s}^{(0)}\right)\right]^{-1} \left(s_y^{(0)}, -s_x^{(0)}, 0 \right)^\intercal,
\end{equation}
where $J(\mathbf{\mathbf{s}^{(0)}})$ is the Jacobi matrix of the system (\ref{eq:sxyz}) calculated at the unperturbed fixed point solution~$\mathbf{s}^{(0)}$. 

In the case of the linearly polarised fixed point $s_+$ with $s_x^{(0)} = -(p+g)$ which is stable below the critical pumping $p_{c}$, the correction (\ref{eq:corr1}) provides a non-zero interaction-driven field $as_z^{(1)} = \delta s^{(0)} \left/ \left(p_c -p\right)\right.$.
The condensate response can thus be characterized by the field enhancement parameter $\chi = as_z^{(1)}/\delta$ and the effective $g$-factor
\begin{equation} \label{eq:geff_triv}
    g_{\rm eff} = 1 + \chi = {s_c \over p_c - p}
\end{equation}
responsible for the  condensate pseudospin precession frequency $ \delta g_{\rm eff}$ according to equations (\ref{eq:sxyz}).

Since $g_{\rm eff}>0$ at $p<p_c$, the induced field is aligned with the external one. 
Note that expression (\ref{eq:geff_triv}) diverges at the critical point 
similarly to the  susceptibility behaviour in Landau theory of second order phase transitions, and linearly scales with the condensate population, as shown in Fig.~\ref{Fig1}(b).
Moreover, as $p$ approaches $p_c$ from above, $p \rightarrow p_c^{+}$, the field enhancement parameter $\chi$, which is now governed by the response of the elliptical states (\ref{eq:s0el}), has a similar asymptotic behavior. 
However, it is important to note that $g_{\rm eff} > 0$ only for the state whose built-in field is aligned with the external one, $s_z^{(0)}/\delta > 0$ (the upper elliptical state in Fig.~\ref{Fig1}(b)). 
For the anti-aligned state, a real Zeeman splitting reduces the total effective field. 
That is why we consider the aligned states only. 
In the supercritical case their response diverges as {$s_c/(2|p-p_c|)$} near $p=p_c$. 
However, in the subcritical regime, dynamically unstable states emerging from the pitchfork exhibit a negative response, $as_z^{(1)}/\delta < 0$, reducing the effective field magnitude, -- see the red dashed lines 
 in Fig.~\ref{Fig1}(c).

Figures \ref{Fig1}(b) and \ref{Fig1}(c) show that close to the critical pumping, polariton pseudospin perceives up to hundredfold increase in the Zeeman field magnitude. 
One of the striking manifestations of this strong field enhancement effect can be observed in the topological properties of Bogoliubov excitations in polariton lattices \cite{Sigurdsson2019,Harrison2023}. 
Since emergence of nontrivial topological phases typically requires time-reversal symmetry breaking, the strong magnetic field is needed -- up to 5~T in Ref.~\cite{Klembt2018}.
However, the elementary excitations emerging in the condensate are also subject to the effective field, created by the condensate pseudospin polarisation and potentially significantly exceeding the external one.


In order to estimate the efficiency of the proposed field-enhancement principle for manipulating topology of bogolons, we proceed with the analysis of the dynamics of single condensate elementary excitations. 
In particular, we focus on the precession of weakly perturbed stationary pseudospin in the presence of finite magnetic field $\delta$.  

Dynamics of the perturbation $\bm{\mathfrak{s}}$ of the fixed point state $\mathbf{s}$ is governed by the linearized system (\ref{eq:sxyz}), i.e. its Jacobi matrix, $\dot{\bm{\mathfrak{s}}} = J(\mathbf{s}) \bm{\mathfrak{s}}$. Therefore, its naturally expected that the excitation experiences the same effective field as the pseudospin state, $g_{\rm eff}\delta = as_z^{(0)} + \delta$.
However, due to the intrinsic non-Hermiticity of $J$, the evolution of the weak perturbation strongly differs from the corresponding pseudospin dynamics.
  
The $\bm{\mathfrak{s}}(t)$-evolution can be given in terms of eigenvalues $\lambda_i$ and eigenvectors $\mathbf{v}_i$ of $J(\mathbf{s})$. 
Since the Jacobian of (\ref{eq:sxyz}) is purely real, it has either three real or a single real and two complex conjugate eigenvalues. 
We are interested in the latter case where the perturbation dynamics reads:
 \begin{eqnarray}\label{EqnDiff} \label{eq:BogDyn}
     \bm{\sigma}(t) = C_1 e^{-\gamma_p t}\left[\mathbf{u}_1 \cos(\Omega t + \varphi) -\right. \nonumber \\ 
     \left. \mathbf{u}_2  \sin(\Omega t + \varphi) \right]  + C_2 \mathbf{v}_3 e^{-\Gamma_p t}.
 \end{eqnarray}
Here $\mathbf{v}_1 = \mathbf{u}_1 + i \mathbf{u}_2$ is the eigenvector corresponding to either of complex conjugate eigenvalues $\lambda_{1,2} = -\gamma_p \pm i\Omega$, $\mathbf{v}_3$ is the eigenvector for purely real $\lambda_3 = -\Gamma_p$.
Coefficients $C_{1,2}$ and $\varphi$ are real.

\begin{figure}
    \centering
    \includegraphics[width=0.95\linewidth]{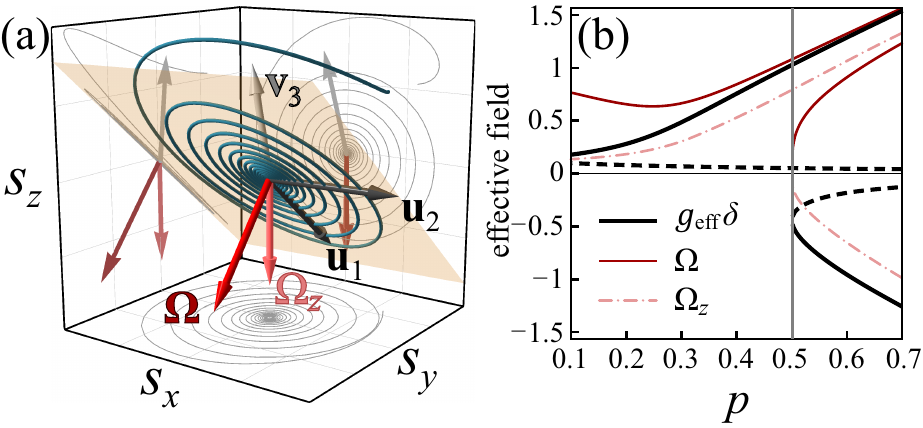}
    \caption{\textbf{Fig. 2.} 
    (a) A schematic evolution of the Bogoliubov's elementary excitation $\bm{\mathfrak{s}}(t)$ shown with the blue line in the pseudospin space. The end-point to which the spiral curls up corresponds to the dynamically stable stationary state $\mathbf{s}$. The pale orange plane contains vectors $\mathbf{u}_{1,2}$ while the vector $\bm{\Omega}$ is its normal. 
    (b) Comparison between definitions of the effective magnetic field acting on the Bogoliubov's elementary excitation no top of the stable pseudospin states at $\delta=0.1$, $a=2.28$ and $g=0.1$. The $p$-dependencies of the precession frequency $\Omega$ and the effective magnetic field ${\Omega}_z$ illustrated in the panel (a).  The vertical gray line indicates position of the fold bifurcation, -- cf. with Fig.~\ref{Fig1}(c). 
    }
    \label{Fig2}
\end{figure}

The long-term evolution (\ref{eq:BogDyn}) of a weak excitation shown in Fig.~\ref{Fig2}(a) represents the dumped rotation with the frequency $\Omega$ in the plane determined by $\mathbf{u}_{1,2}$. 
The normal $\mathbf{n} \perp (\mathbf{u}_1, \mathbf{u}_2)$ assigns direction to the effective magnetic field $\mathbf{\Omega} = \mathbf{n}\Omega$ whose $z$-component $\Omega_z$ is responsible for time-reversal symmetry breaking for Bogoliubov's excitations.
The axis of pure relaxation $\mathbf{v}_3$ is misaligned with $\mathbf{\Omega}$ in contrast to the problem of a classical spin in magnetic field.

In general, $\mathbf{\Omega}$ differs from the effective field $g_{\rm eff}\delta$ acting on a pseudospin itself.
However, in the limit of strong pumping, the excitation dynamics is dominated by the interaction-driven field $a s_z^{(0)}$ that implies $\Omega_z \approx g_{\rm eff}\delta$. 
In contrast, in the vicinity of the bifurcation point, where the giant enhancement of the external field occurs, this simple asymptotics fails.

The values of $\Omega_z$ and $g_{\rm eff}\delta$ near the fold (saddle-node) bifurcation are compared in Fig.~\ref{Fig2}(b). 
Despite significant difference in the involved dynamics, the magnitudes of the effective field for a pseudospin $\mathbf{s}$ and for Bogoliubov's excitations $\bm{\mathfrak{s}}$ follow the same trend.
In addition, the existing mismatch quickly vanishes with the increase of the real field amplitude $\delta$.

In the case of coupled condensate lattices, the topological gap is expected to open in the Bogoliubov's excitation spectrum in the presence of symmetry breaking and spin-anisotropic interactions \cite{Sigurdsson2019}.
Moreover, in the effective field approximation, the uniform spin polarisation is equivalent to a conservative polariton Zeeman field and reduces to Hermitian spin-splitting diagonal terms in the Bogoliubov matrix.
These terms responsible for topological gap opening are proportional to both the uniform spin polarisation and the strength of spin-anisotropic interaction, rendering the corresponding effective Zeeman field identical to the self-induced field $as_z$ in Eqs. \eqref{eq:sxyz}.

\begin{figure}
    \centering
    \includegraphics[width=\linewidth]{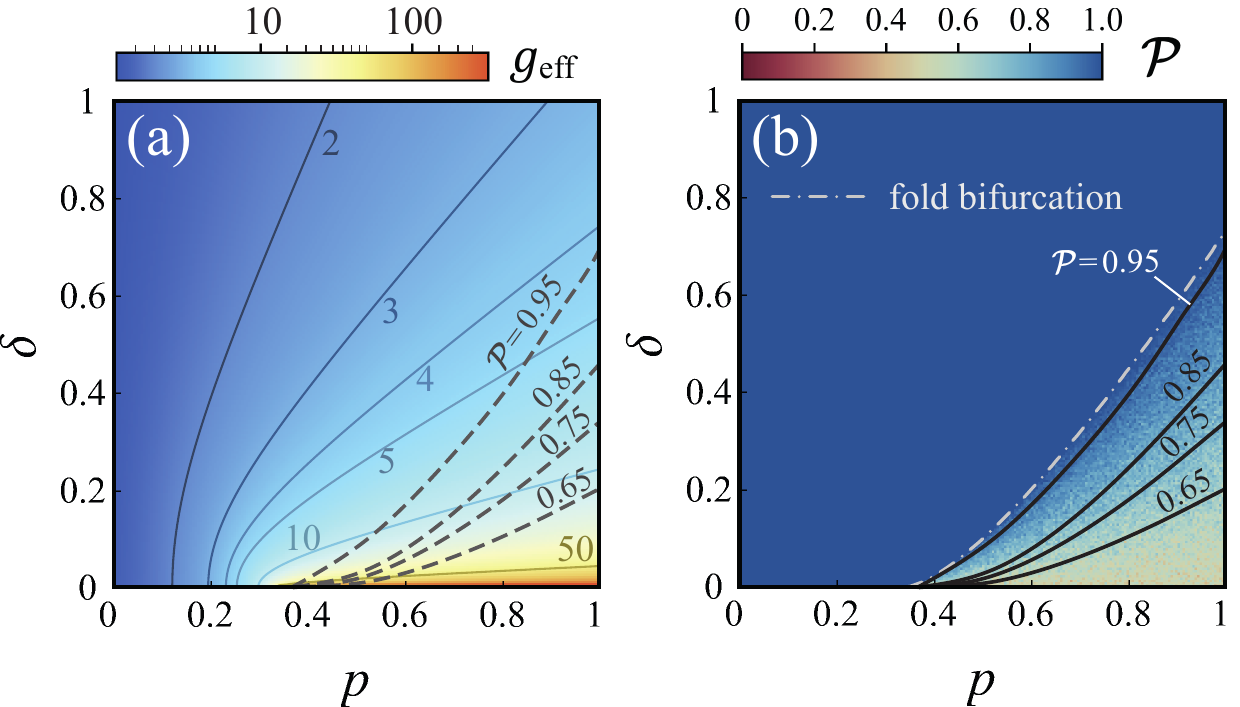}
    \caption{\textbf{Fig. 3.}
    (a) Effective $g$-factor as a function of the pump $p$ and the external magnetic field strength $\delta$. It is assumed that the condensate occupies a stable pseudospin state with $as_z/\delta > 0$. 
    (b) The probability $\mathcal{P}$ of excitation of the state whose interaction-driven field is co-directed with the external one. A bistable regime with $\mathcal{P}<1$ occurs in the region on the right bottom corner of the ($p$,$\delta$) parameter plane. The dash-dotted line indicates position of the fold bifurcation which gives birth to the stable state with anti-aligned effective field. The descending ladder of probability cuts are shown with gray lines which are duplicated in panel (a). Each data point was obtained by a direct numerical solution of (\ref{eq:sxyz}) with random initial conditions and averaging over 200 realizations.
    }
    \label{Fig3}
\end{figure}

In order to find the field-enhancement strength beyond perturbative approach, we numerically search for stable stationary solutions of system~(\ref{eq:sxyz}). 
The resulting $g$-factor $g_{\rm eff} = \left. \left( a s_z + \delta\right) \right/ \delta$ is shown in Fig.~\ref{Fig3}(a) on the $(\delta,p)$ parameter plane. 
Even away from the region of fast divergence, polariton condensate is able to amplify external magnetic field up to an order of magnitude.

The best results can be obtained at the weak magnetic field where $g_{\rm eff}$ diverges according to the perturbation theory. 
This regime, however, has a significant drawback: at strong pumping which favours large $g_{\rm eff}$, polariton condensate exhibits bistability with two stable pseudospin configurations with opposite directions of the interaction-driven effective field. 
Under non-resonant excitation, the building-up of internal magnetization in the spin-bifurcation event occurs spontaneously with randomly selected direction. 
In particular, in the field-free case $\delta \rightarrow 0$, the  condensate excited above the critical pumping $p_c$ occupies either of two elliptically polarised states (\ref{eq:s0el}) with equal probabilities \cite{Ohadi2015}.

However, in the broken symmetry case of finite $\delta$, the balance between spin-up and spin-down states shifts towards the state with the aligned built-in magnetic field corresponding to positive $g_{\rm eff}$. 
In the configuration shown in Fig.~\ref{Fig1}(c), it is the upper state with $s_z>0$.

The probability $\mathcal{P}$ of occupation of the aligned state is shown in Fig.~\ref{Fig3}(b) in the parameter space spanned by $p$ and $\delta$.
Within the domain of $\mathcal{P} = 1$, the condensate supports a single stable configuration. 
Far away from the low-$\delta$ limit, this regime provides quite small $g_{\rm eff}$ about few units, -- see Fig.~\ref{Fig3}(a). 
Above the critical pumping (dash-dotted line) corresponding to the fold bifurcation shown in Fig.~\ref{Fig1}(c), the probability is continuously declining and converges to $\mathcal{P}=0.5$ as $\delta \rightarrow 0$ or $p \rightarrow \infty$.  
Therefore, the optimal regime, where strong field enhancement is combined with the high predictability of the condensate magnetization direction, is reached in a close vicinity to the bifurcation conditions.

In conclusion, we summarize the obtained results. 
Intrinsically non-equilibrium bosonic condensate of exciton polaritons can spontaneously develop strong effective magnetic fields due to the spin bifurcation phenomenon \cite{Ohadi2015}.
Since the very first studies \cite{Shelykh2004,Laussy2006} this field was associated with the so-called self-induced Larmor precession of the Stokes vector of the light emitted by the condensate. 
Here we demonstrated that this phenomenon is controllable with an external permanent magnetic field and has a pronounced manifestation at the level of Bogoliubov's excitations {\cite{Sigurdsson2019}}.
{The nonequilibrium condensate exhibits a very sensitive response to the applied magnetic field providing its strong enhancement. }
The enhancement strength can be characterised  by the effective $g$-factor which exceeds one hundred and strongly depends on the pump intensity and the external field magnitude.

The probabilistic character of the steady condensate spin polarisation and thus the direction of the effective field reduces the available parameter range down to the region near the fold bifurcation of the condensate pseudospin.
{Outside of this range, in particular, where spin multistability takes place, polariton spin fluctuations are expected to destabilize the condensate at sufficiently high temperatures \cite{Glazov2013, Ryzhov2016}.}

In addition, we notice a peculiar connection with the Landau theory of second-order phase transitions.
The magnetic-field susceptibility of the condensate diverges near the critical point where a circular polarisation $s_z \neq 0$ appears spontaneously.
For the symmetry-broken states, the divergence at $p=p_c$ is twice slower.

The obtained results pave the way to further investigations of the topological properties of Bogoliubov's excitations in the lattices of driven-dissipative polariton condensates.

\textit{Acknowledgements.} The work of I.C. (analysis of the non-Hermitian dynamics, text writing) is supported by the Russian Science Foundation Grant No. 22-72-00061. The work of A.N. (analytical calculations, work supervision, text writing) is supported by he Russian Science Foundation under Grant No. 22-12-00144.

\end{document}